\documentclass[10pt, twocolumn]{article} 

\usepackage{amsmath,amssymb,graphicx}

\begin{document}

\newcommand{\etal}{\MakeLowercase{\textit{et al. }}} 
\long\def\comment#1{} 
\long\def\fix#1{***Begin FIX--- {\bf #1 } ---end FIX***}
\def\missref#1{{\bf [MISSREF--#1] }}
\long\def\fixref#1{[REF--{\bf #1}--FIX]}
\newcommand{\Argmax}[1]{{\operatorname{argmax}}_{#1}\;}
\newcommand{\Argmin}[1]{{\operatorname{argmin}}_{#1}\;}
\newcommand{\E}{\mathop{\mathbb E}}
\newcommand{\Hop}{\mathop{\mathcal H}}
\newcommand{\Wop}{\mathop{\mathcal W}}
\newcommand{\Rop}{\mathop{\mathcal R}}

\newcommand{\argmax}[1]{\underset{#1}{\operatorname{argmax}}\;}
\newcommand{\argmin}[1]{\underset{#1}{\operatorname{argmin}}\;}

\long\def\ie{\emph{i.e.}} 
\long\def\ai{\emph{a.i.}} 
\long\def\eg{\emph{e.g.}} 
\long\def\wrt{\emph{w.r.t.}} 
\long\def\define{\overset{def}{=}}
\long\def\one{\mathbf{1}}
\long\def\zero{\mathbf{0}}
\long\def\neigh{\mathcal{N}}
\long\def\Lcal{\mathcal{L}}

\title{\bf Two step robust fringe analysis method \\ with random shift}

\author{Mariano Rivera$^1$, Oscar Dalmau$^1$, Adonai Gonzalez$^2$, Francisco Hernandez$^1$ \\ 
\\
$^1$Centro de Investigacion en Matematicas AC, 36240, Guanajuato Gto., Mexico\\
$^2$Centro de Investigaciones en Optica AC, 37000, Leon Gto., Mexico}

\maketitle 

\begin{abstract} 
We propose a two steps fringe analysis method assuming random phase step and changes in the illumination conditions. Our method constructs on a Gabor Filter--Bank (GFB)  that independently estimates the phase from the fringe patterns and filters noise. As result of the GFB we obtain the two phase maps except by a random sign map. We show that such a random sign map is common to the independently computed phases and can be estimated from the residual between the phases. We estimate the final phase with a robust unwrapping procedure that interpolates unreliable phase regions. We present numerical experiments with synthetic and real date that demonstrate our method performance.
\end{abstract}

\section{Method}
In recent years there has been an interest for developing two--steps algorithms with random step; see for example the methods in \cite{Deng20124669, Vargas:12, Ma2014205, trusiak:2015} and references therein. Those techniques have significantly reduced the acquisition time and have simplified the experimental setups. In this work we propose a robust algorithm that can overcome the limitation of random two-steps algorithms for dealing with variable illumination condition and noise. The proposed method is able to estimate the phase from two noise fringe pattern (FP) with a random phase step between them including  temporal variations in illumination conditions and noise. We proceed as follows: first we motivate our algorithm and then we present the details. 

In this work we assume the following FPs models
\begin{eqnarray}
	I_1(x) &=& a_1(x) + b_1(x) \cos(\phi(x) ) + \eta_1(x),  \label{eq:I1} \\
	I_2(x) &=& a_2(x) + b_2(x) \cos(\phi(x) + \delta ) + \eta_2(x); \label{eq:I2}
\end{eqnarray}
where $ x=[x_1, x_2 ]^ \top$ denotes the pixel position in a regular lattice $\Lcal$. The unknowns are: the background illumination, $a_1$ and $a_2$;  the local fringe contrast, $b_1$ and $b_2$;  the independent noise, $\eta_1$ and $\eta_2$; the phase map we are interested in computing, $\phi$; and the random shift between the FPs, $\delta \in (-\pi, \pi)$. In this work, we consider the standard assumptions used in single FP algorithms: $a_i$, $b_i$ (for $i=1,2$), and $\phi$ are smooth.  

\begin{figure}[ht]
\includegraphics[width=\linewidth]{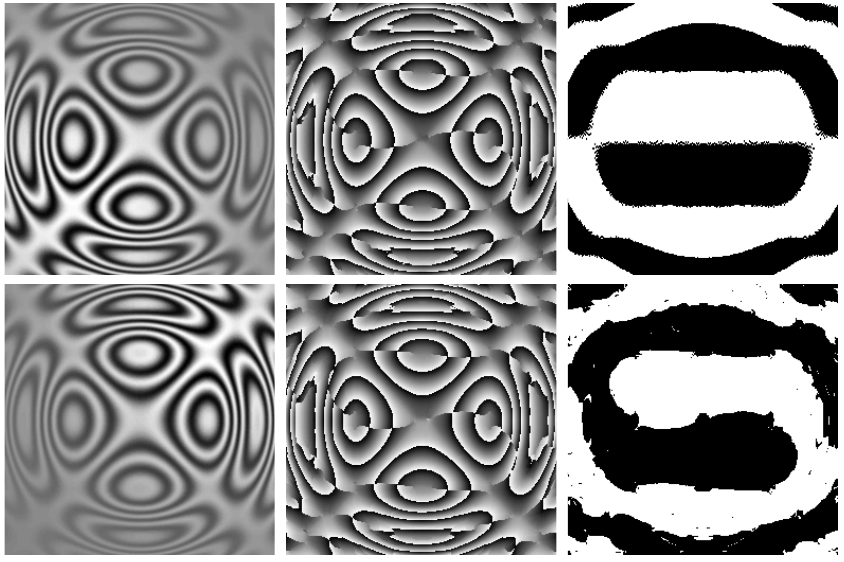}
\caption{First column: two fringe pattern with a random phase shift ($\delta=9\pi/10$). Second column: computed phases with the operator $\Hop$. Third column: sign--map of the actual-phase derivatives by column and estimated sign--map using the proposed method.}
\label{fig:scheme}
\end{figure}
Essentially the proposed method consists of three stage. Firstly, estimation of the wrapped phase with a sign ambiguity, this os obtain by using the $\Hop$ operator. Second, estimation of the correct sign map based on the operator $\Rop$ that computes the wrapped residual phase between a pair of phase map. Finally, we present a robust unwrap process, denoted by the operator ${\Wop}^{-1}$, that interpolates unreliable estimated phase pixels. Following we define the method and the operators. After that, we provide the details. 

Assume we have a method, represented for the operator $\Hop$, such that:
\begin{eqnarray}
	\psi_1(x) \define \Hop \{ I_1 \}(x) &=& \Wop \{ s(x) \phi(x) \} \label{eq:H1}, \\
	\psi_2(x) \define \Hop \{ I_2 \}(x) &=& \Wop \{ s(x) (\phi(x)+ \delta) \}  \label{eq:H2};
\end{eqnarray}
where $s \in \{ -1, 1\}^{\sharp \Lcal}$ is a random sign field and $\Wop$ is the phase wrapping operator, see \cite{riveraARM}.
Next, we use the operator $\Rop$ that computed the residual wrapped phase between two given phases $\psi_1$ and $\psi_2$. This operator is denoted by 
\begin{equation}
\label{eq:residualphase}
	 \Rop(\psi_2,\psi_1) \overset{def}{=} \Wop \{ {\Wop}^{-1} \{\psi_2\} - {\Wop}^{-1} \{\psi_1 \}\},
\end{equation}
where $\Wop^{-1}$ is the unwrapping operator, see \cite{riveraARM}. Since $s(x) \delta \in (-\pi, \pi)$, we obtain $\Rop(\psi_2(x), \psi_1(x)) = s(x) \delta$. Thus, we can estimate the sign with
\begin{equation}
	\tilde s(x) = \left\{   
	\begin{matrix}
		1  &  \Rop(\psi_2(x), \psi_1(x)) \ge 0 \\
      		-1 & \mbox{otherwise} \\
   	\end{matrix}
	\right.
\end{equation}
Let $g(x) = \tilde s(x) \psi_1(x)$ be the wrapped computed phase; then, we can finally estimate the phase $\phi$  with
\begin{equation}
	\tilde \phi(x) = {\Wop}^{-1} \{g(x)\}.
\end{equation}
Essentially, this is the proposed method. Now we specify each one of the used operators: phase analysis ($\Hop$), residual phase ($\Rop$) and unwrapping (${\Wop}^{-1}$). 

\begin{figure}[t]
\includegraphics[width=\linewidth]{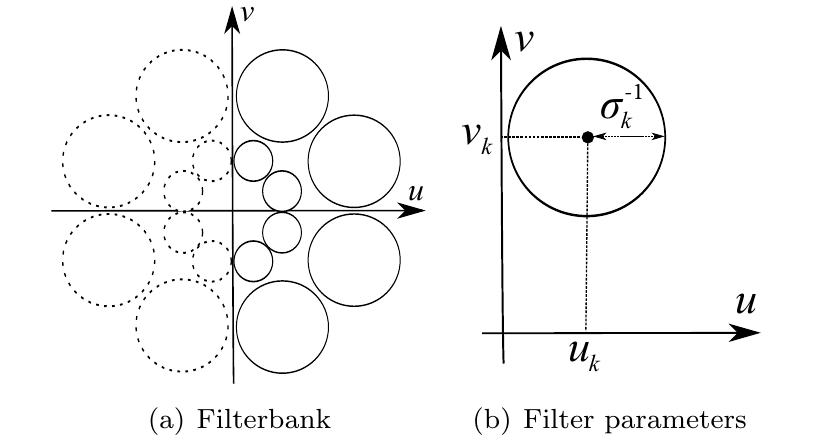}
\caption{Gabor filterbank. a) Filterbank allocated at the half--frequency space (solid--line circles). b) Parameters of a filter in the bank: $\omega_k = [u_k, v_k]^\top$ and $\sigma_k$.}
\label{fig:gabor}
\end{figure}

{ Robust Phase Analysis}, $\Hop$. 
In order to compute the local phase from a noisy FP with variation in the background illumination and contrast, and closed fringes, one can use a general closed fringe method which simultaneously estimates the phase and the sign. However,  those methods are computationally expensive. Therefore, as we have stated, our proposal takes advantage of the second shifted--FP for estimating the sign. Then we propose to use a Gabor's Filter-Bank (GFB); \ie, a set of narrowband filters that only cover a half of the Discrete Fourier Domain and rejects: low-pass region related with the background illumination (DC) spectra and very-high frequencies assumed noise. 
Gabor Filters (GF) are bandpass filters  which are obtained by modulating a sinusoid with a Gaussian \cite{Daugman}. The complex form of the convolution kernel is
\begin{equation}
	\label{eq:gabor}
	h_k(x) = f_k(x) c_k(x)
\end{equation}
where
\begin{equation}
	f_k(x) =  \exp \left[  - {x^\intercal x}/({2\sigma^2_k}) \right]
\end{equation}
\begin{equation}
	c_k(x) =  \exp \left[  -  i \omega_k ^\intercal x \right]
\end{equation}
with $i = \sqrt{-1}$, $\sigma_k$ the width of the Gaussian filter (bandwidth of the bandpass filter) and $\omega_k= [u_k, v_k]^\top$ the central complex frequency (center of the bandpass filter). The GF \eqref{eq:gabor} can be understood by transforming it to the Fourier space. The transform of the $c_k$ term corresponds to a Dirac delta centred at $\omega_k$, $\delta(\omega-\omega_k)$, and the transform of the Gaussian $g_k$ is another Gaussian $G_k$. Then, by the convolution theorem of the Fourier transform, the Gabor filter in the frequency domain is given by $H_k(\omega) = G_k(\omega-\omega_k)$; see Figure \ref{fig:gabor}.

\begin{figure}[t]
\includegraphics[width=\linewidth]{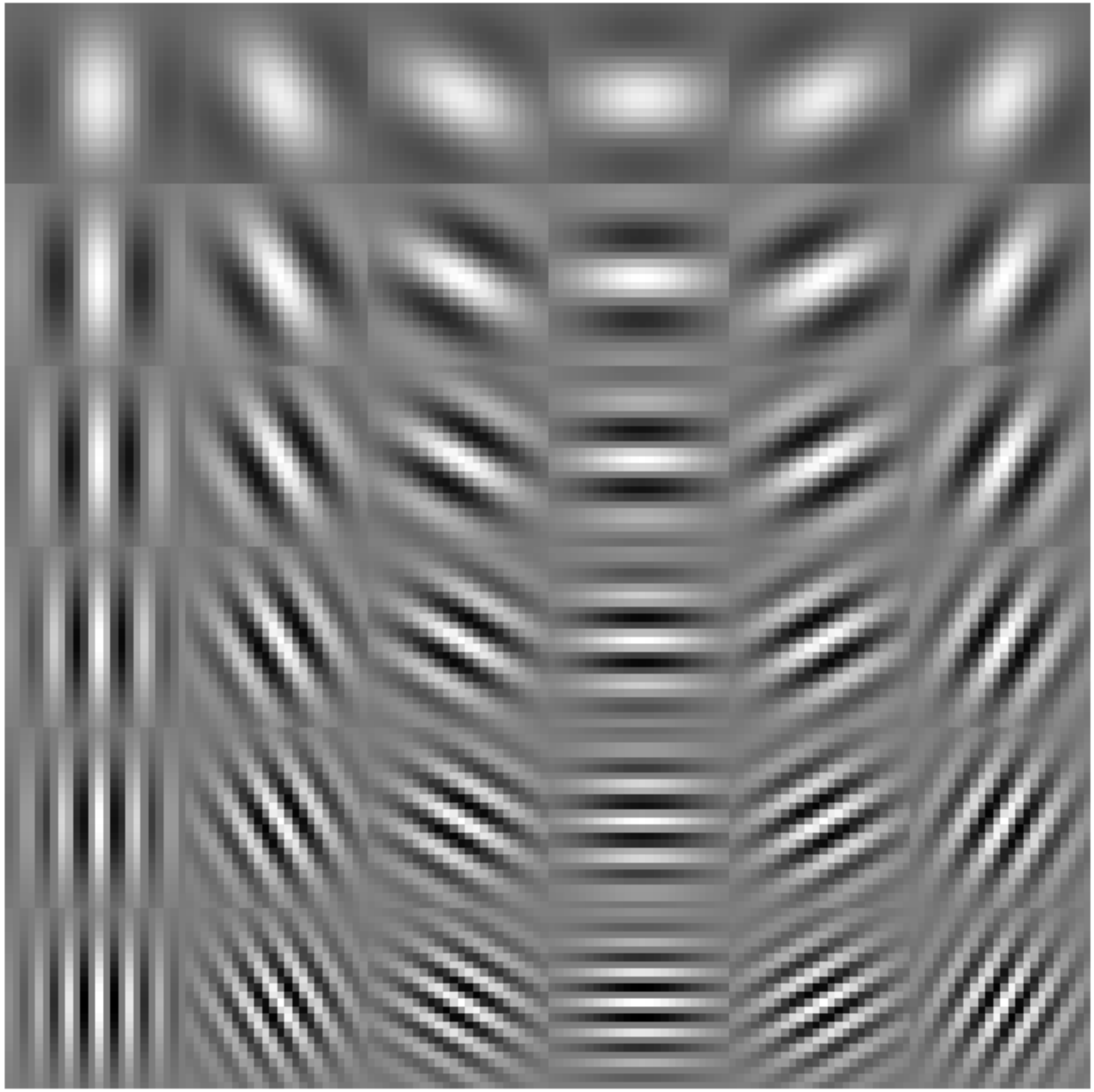}
\caption{Real component of the Gabor's kernels in the spatial domain: kernel size equal $n \times n$ pixels; with $n=24$. The used parameters of the filterbank, expressed in polar coordinates, are  $| \omega_k | = 2\pi k/n $ and $\arg(\omega_k) = \pi k/6$, with  $k = 1,2, \ldots,6$. }
\label{fig:filterbank}
\end{figure}

Hence, the result of applying the $k$th GF to the $j$th fringe pattern is given by:
\begin{equation}
	\tilde I_j^{k} =  h_k \ast I_j \;\; \mbox{for\;\;} j=1,2; k =1,2,\ldots, K;
\end{equation}
where $\ast$ denotes the convolution. $\tilde I_j^{k} (x)$ is complex and expressed in rectangular coordinates (real and imaginary parts), the respective polar coordinates (magnitude and phase) are computed with  
\begin{eqnarray}
	m_j^{k}(x) &=& |  \tilde I_j^{k} (x)|, \label{eq:mik}\\
	\psi_j^{k}(x) &=& \arg(  \tilde I_j^{k} (x)), \label{eq:phiik}
\end{eqnarray}
for $ x \in \mathcal{L}$. Since the FP has locally a dominant frequency, in order to estimate the phase and magnitude at each pixel, we detect the filter with maximum response (the best tuned filter to the local frequency of the FP):
\begin{equation}
	k_j^{*}(x) =  \argmax{k} m_j^{k}(x)
\end{equation}
Thus, the magnitude and phase corresponding to the pixel, $x$ in the FP, is
\begin{eqnarray}
	m_j (x)     &=& m{_j^{k_j^*(x)}}(x) , \label{eq:mi} \\
	\psi_j (x)   &=& \psi{_j^{k_j^*(x)}}(x), \label{eq:phii}.
\end{eqnarray}
Since there exists a winner filter at each pixel, $k_j^\ast(x)$, we obtain a phase at each pixel $\psi_j (x)$ even if the pixel in question belongs to a low frequency region (pixels in region with almost constant phase). In such a case, the GFB will be activated by white noise $\eta_i$. To reduce such a noise detection, we use the computed magnitude as a confidence measure of the computed phase. We compute the mask of reliable phase with
\begin{equation}
	\label{eq:mask}
	\nu_x = \Biggl\{    \begin{matrix} 
	      1 & m_1(x) > \theta \;\mbox{and}\; m_2(x) > \theta \\
	      0 & \mbox{otherwise}\\
	   \end{matrix} \Biggr.
\end{equation}
where the threshold $\theta$ is a parameter of the method. In our experiments we use a single set of parameters for the GFB: such parameters are illustrated in Fig. \ref{fig:filterbank}.

{Phase differences}, $\Rop$. The next step is to implement \eqref{eq:residualphase} for computing the phase differences between the two estimated phases $\psi_1 (x)$ and $\psi_2 (x)$. According to \cite{riveraARM}, the wrapped residual between two phases can be computed without explicit knowledge of $\Wop ^{-1}$ with
\begin{equation}
\label{eq:residualphaseformula}
	\Rop (\psi_2, \psi_1) = {\rm atan2} \left(  \sin(\psi_2-\psi_1), \cos(\psi_2-\psi_1) \right);
\end{equation}
using the identities:
\begin{equation}
\begin{split}
\sin(\psi_2-\psi_1)  = &   \sin \psi_2 \cos \psi_1 - \cos \psi_2 \sin \psi_1 , \\
\cos(\psi_2-\psi_1) = & \cos \psi_2 \cos \psi_1 +  \sin \psi_2 \sin \psi_1.
\end{split}
\label{eq:identities}
\end{equation}

\begin{figure}[t]
\includegraphics[width=\linewidth]{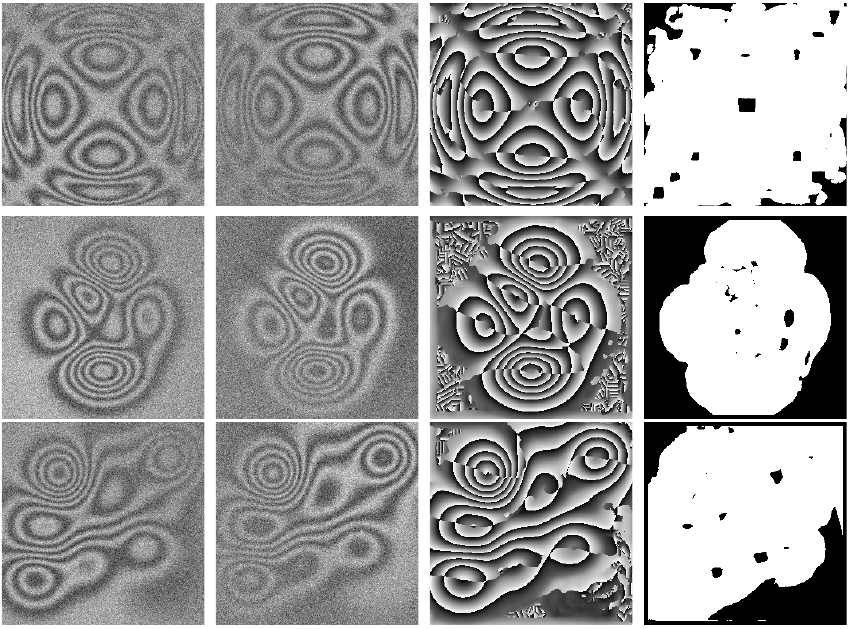}
\caption{First two columns: test FP with $\delta = \pi/10$, $\sigma = 0.5$. Third column: phase $\phi_1$ computed with the Gabor filterbank. Fourth column: region $\nu$  with reliable phase.}
\label{fig:test}
\end{figure}
\begin{figure}[t]
\includegraphics[width=\linewidth]{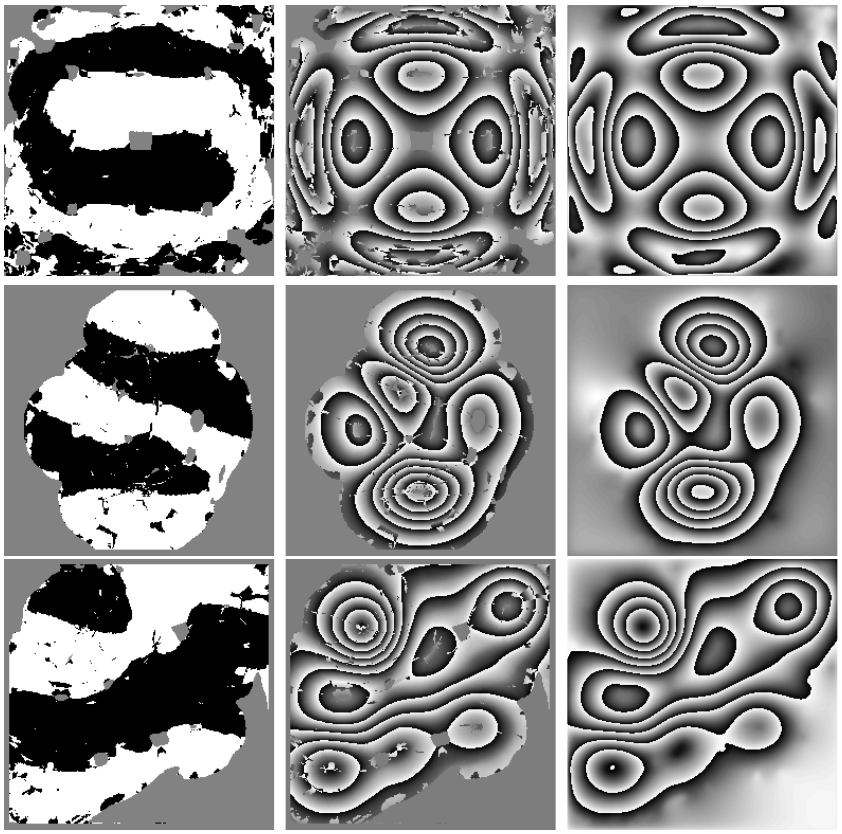}
\caption{First column: estimated sign map $\tilde s$. Second column: Wrapped phase with corrected sign, $g$, in the reliable region.  Third column: rewrapped of the estimated unwrapped phase $\tilde\psi$.}
\label{fig:experiment}
\end{figure}

{Robust phase unwrapping}, $\Wop^{-1}$. The last step is to unwrap the phase taking into account that the computed sign--map $\tilde s$ can be corrupted by errors: compare sign--maps in third column of Fig. \ref{fig:scheme} and experimental results in Fig. \ref{fig:experiment}. We note that the sign is prone to be incorrectly computed at regions where the magnitude of the phase gradients are small (regions with almost constant phase). Such regions can be detected with the mask $\nu$ defined in \eqref{eq:mask}.

In this work we use a variant of the unwrapping algorithm recently reported in \cite{riveraARM}.  The algorithm iteratively update the current estimate $\phi^{(t)}$ of the unwrapped phase  with an unwrapped update phase $\delta^{(t)}$:
\begin{equation}
	\label{eq:delta}
	\phi^{(t+1)}(x) = \phi^{(t)}(x) + \delta^{(t)}(x),
\end{equation}
where the updating field $\delta^{(t)}$ is computed by  
\begin{multline}
\label{eq:energy1}
 \{\delta ^{(k)}, \omega ^\ast\} = \argmin{\delta, \omega} U(\delta, \omega; \hat \rho^{(t)}, \nu_x, \nu_y)  = \\
 	\sum_{x \in \mathcal{L}}  \sum_{y \in \neigh_x}  \Bigl\{ \omega_{xy}^2  \left[ \nu_x \nu_y  (\delta(x)-\delta(y) -\hat \rho^{(t)}_{xy} )^2 \right. \\
	\left.+   \lambda  ( \delta(x)-\delta(y) )^2 \right] + \mu (1-\omega_{xy})^2   \Bigr\}
\end{multline} 
where $\lambda$ and $\mu$ are positive parameters and we define
\begin{equation}
	\label{eq:rho}
	\hat \rho^{(t)}_{xy} \define \Wop \{ g(x) -  g(y)\}- \phi^{(t)}(x) + \phi^{(t)}(y),
\end{equation} 
 and the set of first neighbour pixels to the pixel $x$ as
\begin{equation}
	\label{eq:Nx}
	\neigh_x \define \{y \in \mathcal{L} : \|x-y\|_2 =1 \}.
\end{equation}
We modified the cost function \eqref{eq:energy1} by introducing the binary weights $\nu_x$ and $\nu_y$ [see \eqref{eq:mask}] that define the region with reliable data. Hence, the phase at regions with invalid data are smoothly interpolated by effect of the regularization.

\begin{figure}
\includegraphics[width=\linewidth]{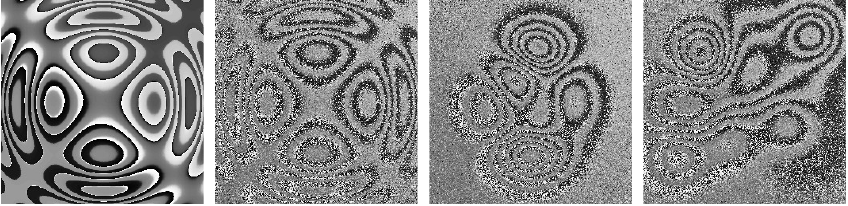}
\caption{Incorrect phases computed with a Gram--Schmidt orthonormalization \cite{Vargas:12}.}
\label{fig:gram}
\end{figure}

\section{Experiments and Conclusions}

Figures \ref{fig:test} and \ref{fig:experiment} demonstrate the proposed method's performance. The test FP are shown in Fig. \ref{fig:test}: we use the same contrast modulation, $b_1$ and $b_2$, as in the FPs of Fig. \ref{fig:scheme}. The GFB robustly  and independently estimate the phase, except by the random sign, of each FP with different background illumination ($a_1 \neq a_2$), contrast ($b_1 \neq b_2$), and corrupted with independent noise ($\eta_1 \neq \eta_2$). An advantage of the GFB is that one can obtain the quality map that defines regions where the estimated phase is reliable. Then, the proposed unwrapping procedure can effectively interpolate the missed phase regions.

Gram--Schmidt based orthonormalization (GSBO) has shown to be a computationally efficient method for computing a couple of quadrature images \cite{Vargas:12}. That interesting proposal has motivated works for overcoming its drawbacks: the performance of such a method is limited when there are variations in the background, contrast and noise; see Fig. \ref{fig:gram}.

\begin{figure}[t]
\includegraphics[width=\linewidth]{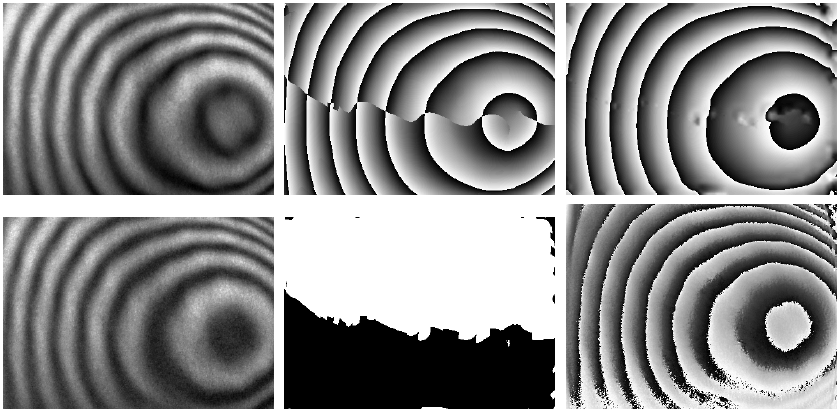}
\caption{Real FP analysis. First column: two fringe pattern with a phase shift equals $\pi/2$. Second column: computed phases with the operator $\Hop$ (top) and computed sign map (bottom). Third column: rewrapped computed phase using the proposed method (top) and computed wrapped phase with GSBO \cite{Vargas:12} (bottom).}
\label{fig:real}
\end{figure}

Fig. \ref{fig:real} shows the coupled phase with the proposed method using as test FP real interferograms. The used parameters of the filterbank, expressed in polar coordinates, are  $|\omega| = 2\pi u/m $ and $\theta = \pi u/n$; with $u = 1,2, \ldots,5$ and  $u = 1,2, \ldots,14$. We can observe that our method correctly recover the phase. In contrast,  we can note that GSBO fails to compute a phase even when the FP have a phase shift equals $\pi/2$. The reason of the GSBO's poor performance is the large changes in the illumination conditions. The GSBO approach can be improved by using a windows-wise technique  \cite{Ma2014205}. Although this  technique  can reduce the effect of variation in illumination components, the main drawback is that it requires a window size that includes  several fringe  fringes, with the additional limitation of processing low frequency FP. In the best of our knowledge, the best procedure for reducing all the mentioned differences is by preprocessing the FPs with banks of quadrature filters (as GFB). In our work, we use the GBF as part of our process. The reader can find limitation of other two-step demodulation algorithms in \cite{Vargas:12, Ma2014205, trusiak:2015}. In \cite{trusiak:2015} it is described a sophisticated preprocess for normalising the fringes in order to apply GSBO. The result of such a prefiltering is similar in our approach to apply the GFB. However, differently to \cite{trusiak:2015} that requires the additional step of GSBO for estimating the phase with the correct sign, in our case, the correction sign map $s$ is obtained directly from the  GFB result; \ie, we do need the extra orthonormalization step.


\end{document}